# A Quantitative Model of Triboelectric Charge Transfer

Karl P. Olson[1], Laurence D. Marks[1]*

[1]Department of Materials Science and Engineering
McCormick School of Engineering and Applied Science
Northwestern University
2220 Campus Drive, Room 2036
Evanston, IL 60208

*Corresponding Author: laurence.marks@gmail.com

## Abstract

Triboelectricity, when rubbing or contacting materials causes electric charge transfer, is ubiquitous across many fields, and has been studied in detail for centuries. Despite this, a complete description of triboelectricity remains elusive. Here, we analyze the contact between a metal asperity and a semiconductor, including contributions from the depletion zone of the semiconductor and from flexoelectric polarization that arises due to the strain gradients at asperity contacts. The free charges involved in charge transfer are then discussed and calculated. As a result, we develop a quantitative model for triboelectric charge transfer that details how charge transfer scales with contact parameters, the relative influence of depletion and flexoelectricity, and which agrees with various trends in multiple classes of triboelectric experiments.

## 1 Introduction

Triboelectricity and contact electrification, the transfer of charge between rubbing and contacting materials, are critical in a wide range of fields and phenomena, from the design of triboelectric nanogenerators [1], accurate dosage of pharmaceutical drugs [2-4], formation of planets [5], pollution of polar snow [6] and enrichment of wheat bran [7] to bee pollination [8], insecticide mechanisms [9], and effective coffee grinding [10]. Despite extensive work on triboelectricity spanning centuries [11-14], there is not yet a complete model of triboelectric charge transfer, especially when one or both contacting bodies are not metals [14].

Early work on triboelectricity focused largely on ordering materials into a "triboelectric series", where materials on one end of the series charged positively when rubbed against materials toward the other end. However, as Shaw found in a series of papers in the early twentieth century, triboelectric series do not explain same-material charging [15], changes of charge transfer sign with pressure [13], triboelectric cycles [16], or the effects of pre-strain [17, 18] and macroscopic bending [13]. Additionally, such a series does not quantitatively predict charge transfer, though there have been some modern efforts to resolve this [19].

Quantitative prediction of charge transfer, up to now, has relied on models with empirical fitting parameters to match experiments. One effective model of contact between two metals is the condenser model, where an analogy to a capacitor is drawn [20, 21]. If the difference between the contact potentials (work functions) of the metals is treated as a potential difference across a capacitor, the charge that compensates the electric field will be the charge transferred. There has been much effort to extend this model to include insulators or semiconductors. However, defining contact potentials for these materials is not straightforward, since surface states which allow states in the band gap to exist are needed to facilitate charge transfer. Furthermore, the choice of the distance between the capacitor plates in the condenser model is arbitrary [22]. Additionally, as we [23-25] and others [26, 27] have shown, electromechanical effects such as flexoelectricity must be included in a complete model. Furthermore, like the triboelectric series, simple condenser models fail to explain some experimental observations such as charge transfer between like materials or changes with macroscopic bending.

Beyond its connection to triboelectricity, flexoelectricity is important in many phenomena at the nanoscale [28]. Flexoelectricity has been used to mechanically pole ferroelectrics [29], create piezoelectric-like composites from centrosymmetric materials [30], and modulate photovoltaics by strain-gradient engineering [31]. Macroscopic measurements of flexoelectric coefficients are typically done by bending beam-shaped samples [32, 33] or compressing truncated pyramidal samples [34]. Recently, there have been efforts to measure flexoelectric coefficients at smaller scales using nanoindentation [35].

Previously, we have shown that triboelectric charge transfer can be driven by contacting asperities [23] and described the potentials and bound charges associated with this contact [25]. Here, we have extended this work to allow quantitative prediction of charge transfer between a metal and a semiconductor, without any empirical parameters. During contact with a metal, charge movement occurs to form a depletion region in the semiconductor. Additionally, bound surface charges form at the surface of a dielectric due to electromechanical effects. Where the materials contact, this bound charge will be compensated by a combination of transferred or redistributed surface charges, ion adsorption, surface reconstruction and relaxation changes, or other phenomena. These compensating charges can be involved in charge transfer. We consider these possibilities, in addition to other factors such as the image effects of the surface and metal. Combining these, we develop a quantitative model of single asperity triboelectric charge transfer that needs no empirical parameters and includes the physics required to explain same material triboelectric charging and the effects of bending and pre-strain.

## 2 Theory

In this section, we develop the analysis for charge transfer in an *ab-initio* sense, where all the terms used have independent theory, have been separately modelled and, in many cases, have been experimentally measured for many materials or material combinations. For most of this work, we consider a rigid spherical $Pt_{0.8}Ir_{0.2}$ indenter of radius $R$ contacting a 0.7 wt% Nb-doped $SrTiO_3$ (Nb:$SrTiO_3$) half-space with a force $F$; later we consider cylindrical and conical indenters, as well as infinitely long cylindrical rollers and square punches. We choose Nb:$SrTiO_3$ because few materials except $SrTiO_3$ have been analyzed in sufficient depth such that quantitative values for necessary terms, such as the flexoelectric coefficients, are known. Note that because $SrTiO_3$ has a centrosymmetric crystal structure, there are no piezoelectric contributions. Metal interfaces with $SrTiO_3$ shows piezoelectric properties [36], but this is an emergent effect from the depletion zone created by the contact with the metal. We analyze the depletion polarization directly, so ignoring piezoelectricity remains valid; adding a piezoelectric term would be double counting.

Before proceeding into details, it is important to clarify the difference between "free charges" and "bound charges", as both terms will be relevant. Free charges are entities such as electrons, holes and ions which can move around in response to an applied potential. Bound charges are a representation of electric field gradients and polarization in terms of effective charges which are fixed in space and not free to move. In the following sections the term "bound" will only be used for these fixed charges, in all other cases we are referring to the free charges.

The analysis is somewhat complete, although there are approximations such as the use of Hertzian contact mechanics [37-40], some assumptions involved in flexoelectric coefficients, and neglect of the higher-order coupling of the elastic strain and electromechanical terms. Some higher-order couplings and other mechanical effects are discussed later.

We use herein an approach that sidesteps some of the complications of calculating the full fields and potentials, and instead focuses upon the charge. We first calculate the charges involved in the formation of the depletion potential in the semiconductor and the bound charges created by flexoelectric polarization. The total charge transfer involves a combination of the depletion potential charge and the free charge that balances the flexoelectric polarization at the interfaces. The contributions to the free carrier charges are, with some assumptions that we will come to, split into three terms:

1. Charge at the interface between the metal and semiconductor which compensates the surface polarization. This is equivalent to the condition $D_z = 0$ with $z$ the spatial coordinate normal to the interface and $D$ the (Maxwellian) electric displacement density vector, and the free carrier charge balances the surface polarization.
2. Charge at the surface away from the metal-semiconductor contact, again for $D_z = 0$. This polarization does not have to be compensated only by free carriers, as will be discussed later.
3. Induced charge on the metal due to any remnant polarization of the surface away from the contact (uncompensated bound charges from 2.).

While we specifically focus on a $Pt_{0.8}Ir_{0.2}/Nb:SrTiO_3$ contact, the structure of this theory is applicable to many cases. This will be discussed in more detail later, but we will note that metal/semiconductor, metal/insulator, and metal/polymer triboelectric contacts are not substantially different. All have band gaps and dopants (either deliberate or accidental), and their band-bending and interface physics are essentially the same, with depletion layers forming and possible Fermi level pinning. The details of current flows may be different, but the principles behind charge transfer are not.

## 2.1 Depletion Region

When a metal and semiconductor contact, charge is transferred to equilibrate the Fermi levels of the materials, forming a space charge in the semiconductor with density $q^s$ and a resulting depletion potential $\psi$. This source of charge transfer is the Volta potential or contact potential that has been incorporated for many decades in modified condenser models [20, 41]. Yamamoto, *et al*, determined the 1-D depletion potential $\psi$ in a Nb:SrTiO$_3$ Schottky diode by solving the Poisson equation with a field-dependent permittivity $\epsilon = \epsilon_0 \gamma / \sqrt{\alpha^2 + E^2}$, where $E$ is the electric field, and $\alpha$ and $\gamma$ are constants fitted to experiments; at room temperature they are 0.04 V/nm and 14.8 V/nm, respectively [42]. When there is no applied bias, the depletion potential $\psi$, electric field $E$, and depletion width $W$ are

$$\psi(z) = -\frac{\alpha}{\beta}\left[\cosh(\beta(W-z)) - 1\right] + \zeta_F \quad (1a)$$

$$E(z) = -\psi'(z) = -\alpha \sinh(\beta(W-z)) \quad (1b)$$

$$W = \frac{1}{\beta}\cosh^{-1}\left(\frac{\beta}{\alpha}V_d + 1\right) \quad (1c)$$

where $z$ is the coordinate distance from the interface, $\beta = eN/\epsilon_0\gamma$, $e$ is the elementary charge, $N$ is the carrier concentration (i.e., Nb dopant concentration, $2.24 \times 10^{26}$ m$^{-3}$ for 0.7 wt% doping), and, when no external bias is applied, $V_d = \Phi_b + \zeta_F$. $\Phi_b$ is the Schottky barrier, approximated in the Schottky-Mott limit by 1.4 eV, the difference between the Pt$_{0.8}$Ir$_{0.2}$ work function of roughly 5.4 eV [25, 43-45] and the SrTiO$_3$ electron affinity of 4 eV [46]. $\zeta_F$ is the Nb:SrTiO$_3$ Fermi level measured with respect to the conduction band and is 65 meV for 0.7 wt% doped Nb:SrTiO$_3$, as calculated with WIEN2k [47].

The space charge density is $q^s(z) = d/dz\,(\epsilon E) = d/dz\left(-\epsilon_0 \gamma \tanh(\beta(W-z))\right) = eN\,\text{sech}^2(\beta(W-z))$. Thus, the total charge transferred during the formation of a depletion region with interface area $A$ is, within a 1D approximation,

$$|\Delta Q_d| = A\int_0^W q^s(z) = A\int_0^W eN\,\text{sech}^2(\beta(W-z))\,dz = \epsilon_0\gamma\tanh(\beta W) = \epsilon_0\gamma\frac{\sqrt{\delta^2-1}}{\delta} \quad (2)$$

where $\delta = \beta V_d/\alpha + 1$.

For 0.7 wt% Nb:SrTiO$_3$, substituting the values given above, this is

$$\frac{\Delta Q_d}{A} = +0.81 \text{ e/nm}^2 \quad (3)$$

The sign above is chosen so that charge transfer (electrons, holes, or ions) has a positive sign for transfer from the n-type semiconductor to the metal (i.e., positive when the metal ends positively charged). This sign convention holds for the remainder of the work.

The above analysis is accurate up to a 1D approximation; it does not consider the edges of contact, where not only the region directly below the contact is depleted, but also short distances outside of the contact. If we assume the 1D solution holds for the distance from the interface instead of the depth $z$, we have instead

$$|\Delta Q_d| = A\int_0^a q^s(z)\,dz + 2\pi\int_a^{a+W}\int_0^{\sqrt{W^2-(\rho-a)^2}} q^s\left(\sqrt{z^2+(\rho-a)^2}\right)dz\,\rho d\rho \quad (4)$$

where $\rho$ is the radial distance from the center of contact and $a$ is the contact radius. For a contact between a Pt$_{0.8}$Ir$_{0.2}$ with radius 60 nm and SrTiO$_3$ at a force of 4 µN, we find $|\Delta Q_d| = 189$ e with the 1D solution only and $|\Delta Q_d| = 1213$ e including the region outside the contact. That is, when the contact radius and depletion width are comparable (here, $W = 11.3$ nm and $a = 8.6$ nm), this increases $\Delta Q_d$ significantly compared to the pure 1D solution. When $a \gg W$, this extra term is small compared to the 1D solution and may be ignored. For example, if $a = 10W$, the 1D solution is 80% of the 3D solution. Because typical experiments have contact radii [20] much larger than the depletion width [48, 49], this extra term is applicable only in select cases. Therefore, we will continue to use the 1D solution $|\Delta Q_d|/A = 0.81$ e/nm$^2$. A more detailed 3D analysis is given in Appendix 1.

In principle, the depletion charge may be coupled to the mechanical deformation in the semiconductor due to the contact. For the highly-doped Nb:SrTiO₃ considered here, there is very little coupling with strain. For low doping, the coupling is stronger, rising up to a factor of $1 + \varepsilon$ where $\varepsilon$ is the strain. Appendix 2 includes more details regarding the coupling. Note that the field-dependent dielectric term used above is an ansatz that allows the interface electrostatics problem to be solved using the Poisson equation without the need to account for the change in available states caused by band bending via the Poisson-Schrödinger equation. While flexoelectric coefficients are related to the dielectric permittivity [28], they are therefore not coupled with this effective field-dependent permittivity.

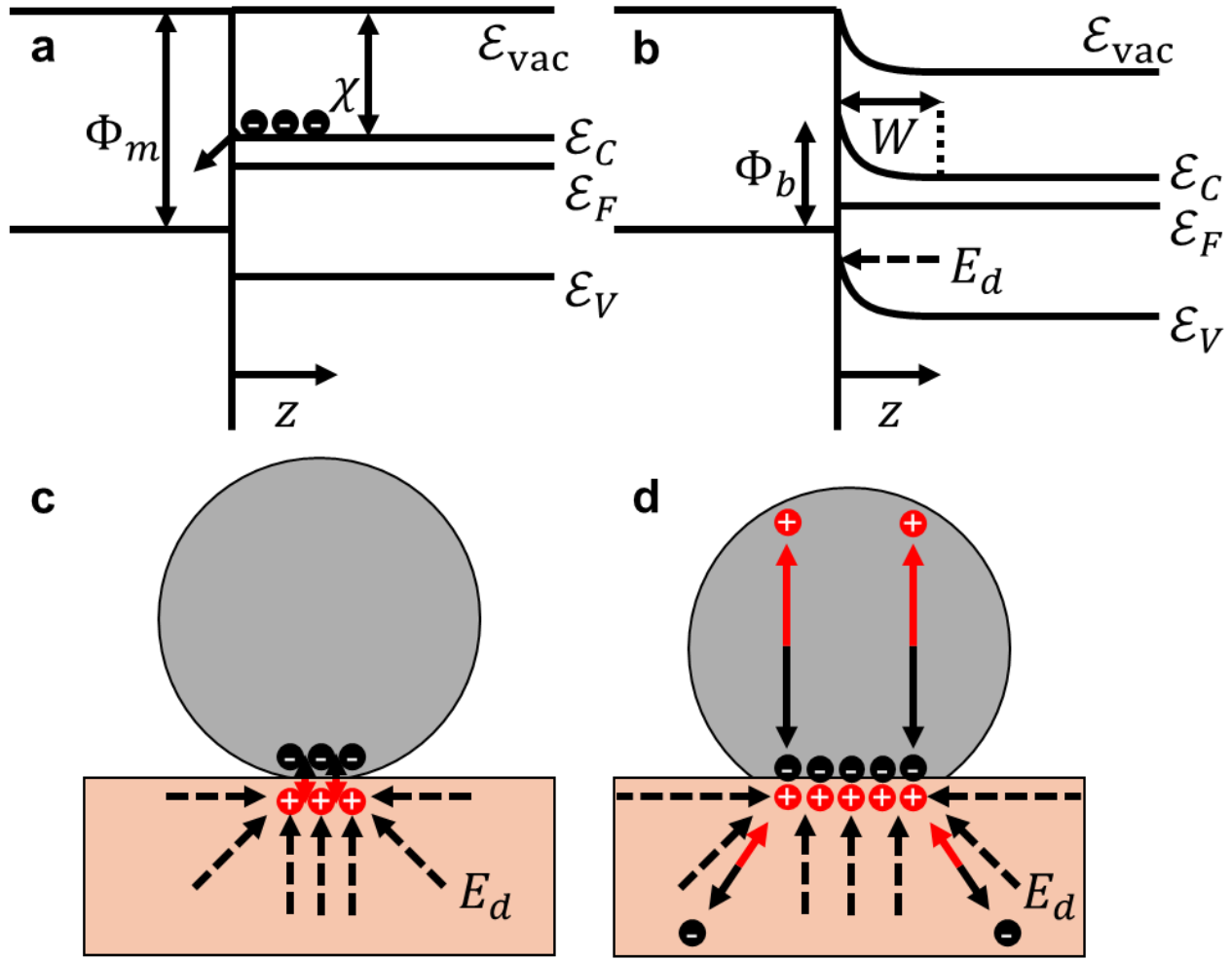

Fig. 1. (a-b) Band diagrams for a Schottky diode (a) just before contact, the flat band case, and (b) with a partially formed depletion region, picoseconds after contact. An electric field $E_d$ develops in the depletion region. $\mathcal{E}_C$, $\mathcal{E}_V$, $\mathcal{E}_F$, and $\mathcal{E}_F$ are the conduction band minimum, valence band maximum, Fermi level, and vacuum level, respectively. $\Phi_m$, $\chi$, and $\Phi_b$ are the metal work function, semiconductor electron affinity, and Schottky barrier height, respectively. $W$ is the depletion width. (c-d) Schematic diagrams of the depletion charge transfer mechanism. (c) At the initial contact, electrons will transfer from the semiconductor to the metal, forming an electric field $E_d$. This is shown in 1D in (a) and (b). (d) Later during the contact, $E_d$ prevents further electron transfer from the semiconductor, so they move to the bulk to form the growing depletion region. To compensate this charge, electrons from the metal

flow to the interface. As contact is released and the depletion region collapses, the electrons at the interface balance the positive space charge and the metal (semiconductor) is left positively (negatively) charged.

The sign of this charge is not as simple as a 1D approximation; the dynamics of contact are likely important. Experiments involving metal/semiconductor contacts, such as metal/semiconductor sliding contacts in direct-current triboelectric nanogenerators [50, 51] and particle contacts of various minerals with multiple metals [52], show that the metal should charge positively for the present set of materials, and that generally, the metal should charge more positively when contact involves metals with higher work functions (i.e., when the depletion charge is larger). In the simple 1D formation of a Schottky barrier, electrons transfer from the semiconductor to the metal, as shown in the flat-band diagram of Fig. 1a. However, note that as the depletion region forms, even incompletely as shown in Fig. 1b, an electric field $E_d$ develops. At the start of contact, electrons will transfer to the metal as the depletion region first begins to form, as shown in Fig. 1c. As the contact proceeds, due to the electric field that has already formed, further formation of the depletion region involves electrons moving into the semiconductor bulk, thereby forming the positive space charge. To compensate the space charge, electrons from the metal will move to the interface. This process is shown in Fig. 1d. Finally, as the materials come out of contact, the electrons at the interface will remain on the semiconductor, leaving the metal positively charged and the semiconductor negatively charged.

Note that the calculation of $\Delta Q_d$ is still valid in the case electrons move into the semiconductor bulk and when there is an inhomogeneous dielectric constant. Noting the form for Poisson's equation with an inhomogeneous dielectric of

$$-\nabla \cdot \big(\epsilon(\mathbf{r})\nabla\psi(\mathbf{r})\big) = \nabla \cdot \big(\epsilon(\mathbf{r})\mathbf{E}(\mathbf{r})\big) = q_s(\mathbf{r}) \tag{5}$$

Integrating from deep in the material for a 1D case gives

$$\epsilon E_d \cdot \hat{z} = \epsilon(0)E(0) = \int_0^W \frac{d}{dz}(\epsilon E)\, dz = \int_0^W q_s\, dz = |\Delta Q_d| \tag{6}$$

because $E(W) = 0$, where, as before, $W$ is the depletion width. Therefore, assuming this process is dominated by the electrons transferring into the semiconductor bulk to form the space charge as in Fig. 1d, then the charge transferred is $+\Delta Q_d$.

The dynamics of contact with respect to depletion region formation support the plausibility of this mechanism. The contact time for a particle with radius $R$ and density $\eta$ impacting a half-space at velocity $v$ is, according to Hertzian mechanics [53],

$$t = 5.08 \left(\frac{\eta}{Y^*}\right)^{2/5} \frac{R}{v^{1/5}} \tag{7}$$

where $1/Y^* = (1-v_1^2)/Y_1 + (1-v_2^2)/Y_2$, with $v_i$ and $Y_i$ the Poisson's ratios and Young's moduli of the two materials, respectively. For an SrTiO$_3$ particle with radius 60 nm impacting a Pt$_{0.8}$Ir$_{0.2}$ half-space at 10 m/s, the contact time is 200 ps [54, 55]. Using rough values of $R = 500$ μm, $Y = 1$ GPa, $v = 0.33$, and $\eta = 1$ g/cm$^3$ for a polymer particle typical in many triboelectric particle impact experiments, the contact time is 6 μs. Meanwhile, the formation time for the depletion region in Schottky contacts is of order picoseconds [56]. Thus, the contact times, even for small, stiff materials, are much longer than the depletion region formation time. Therefore, it is reasonable to suggest the formation of the depletion region early during the contact, as in Fig. 1c, can affect the formation of the depletion region later during the contact, as in Fig. 1d.

## 2.2 Flexoelectricity

There are two components to the flexoelectric contribution: the bulk part and what is often called "surface flexoelectricity" which involves gradients of the shift of the mean inner potential (MIP) with strain [57]. In previous work [25, 58], we separated these to work with the potentials, as those are relevant for experimental measurement of I-V curves. Herein, this is not needed, so we use macroscopic flexoelectric coefficients. The MIP terms are inherently included in the experimental measurements, so we use estimates of the flexoelectric tensor components based on previously published experimental data [59]. Extracting flexoelectric tensor components from experiments necessarily involves some assumptions, as measuring the complete tensor is problematic [33, 60]; more detail is found in Appendix 3. We use $\mu_{iiii} = -380$ nC/m, $\mu_{iijj} = -103$ nC/m, and $\mu_{ijij} = -1.4$ nC/m.

The flexoelectric polarization caused by a strain gradient is

$$P_i = \mu_{klij} \varepsilon_{kl,j} \tag{8}$$

where $P_i$ is the $i^{\text{th}}$ component of the polarization density, $\mu_{klij}$ the flexoelectric coefficients, and $\varepsilon_{kl,j}$ is the gradient of the strain component $\varepsilon_{kl}$ with respect to the $j^{\text{th}}$ coordinate. This polarization is equivalent to bound bulk and surface charge densities $\varrho = -\nabla \cdot P$ and $\varsigma = P \cdot \hat{n}$, respectively, where $\hat{n}$ is a unit vector normal to the surface (for the half space, $\hat{n} = -\hat{z}$). These bound charges are not free to move or transfer as they simply represent the polarization in terms of charge densities. Also, note that there is zero net bound charge, which is simply shown using the divergence theorem. The total bulk charge $\iiint_\Omega -\nabla \cdot P \, d\Omega$ is opposite to the total surface charge $\iint_S P \cdot \hat{n} \, dS$.

## 2.3 Surface Charge Compensation and Free Carrier Charge Transfer

As previewed, we now split the remaining analysis into three possible locations: the interface between the metal and semiconductor, surface of the semiconductor away from the contact, and surface of the metal away from the contact. The free and bound surface charges that may exist at these locations will be subsequently called interface, surface, and metal charges, respectively. Note that "surface" will refer specifically to the semiconductor surface. These three locations are shown schematically in Fig. 2.

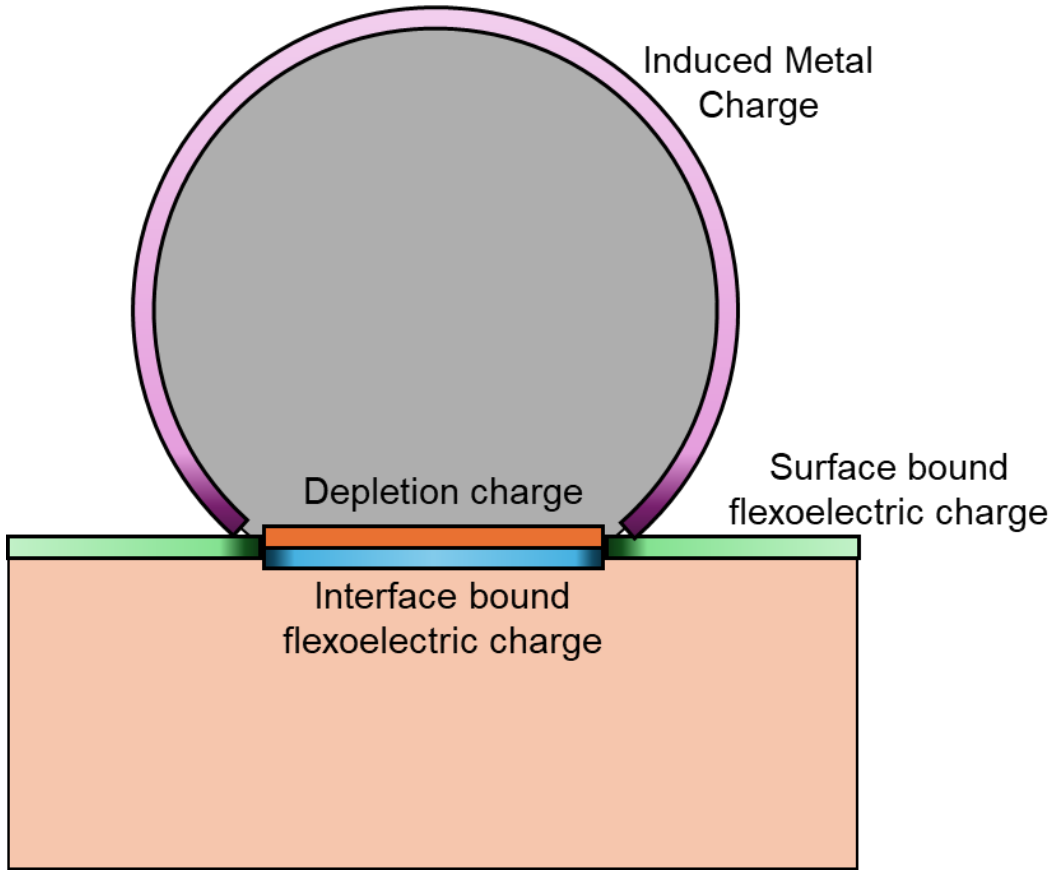

Fig. 2. The depletion charge is on the metal at the interface, with the space charge beneath the contact in the semiconductor. For clarity, the space charge is not shown here. Flexoelectric polarization in the semiconductor results in bound charges at the interface, where the metal and semiconductor contact, and at the surface, outside of the contact. When the surface bound charges are not fully compensated, they induce charges on the metal.

Possible sources of free charges are discussed for each of these locations in the following sections. The specific mechanisms and whether a particular location is involved in charge transfer depend on the materials involved. We will discuss possibilities in general, as well as some concrete points specific to metal/semiconductor contacts.

### 2.3.1 Interface Depletion Charge

At the metal-semiconductor interface, the depletion charge $\Delta Q_d$ will be in interface states mainly in the first nanometer of the metal, depending upon detail such as the electron screening length. They can be considered as compensating for an effective bound charge at the surface as mentioned in Eq. (5), this result being independent of where they come from.

### 2.3.2 Interface Flexoelectric Charge

At the metal-semiconductor interface there will also be free carriers in the metal to compensate the flexoelectric polarization bound charge, which is given by an integral over the contact radius:

$$\Delta Q_{if} = 2\pi \int_0^a -\varsigma(\rho)\, \rho d\rho \qquad (9)$$

Where these free carriers come from will be considered later.

### 2.3.3 Surface Flexoelectric Charge

As at the interface, a bound polarization charge $\varsigma$ exists at the surface. Unlike at the interface, the surface has no metal in contact and therefore these may remain without compensation or be compensated by charges originating from either the semiconductor bulk or the metal and conduction along the surface. Additionally, the bound charges can be compensated by other means, such as by adsorption of atmospheric ions or changes in the surface reconstruction that balance the polarization [61, 62]. These latter cases clearly do not contribute to charge transfer, but the former compensation by free charges can.

Similar to the interface, the total compensating free surface charge is

$$\Delta Q_{sf} = 2\pi \int_a^\infty -\varsigma(\rho)\, \rho d\rho \tag{10}$$

### 2.3.4 Induced Metal Image Charge

When a grounded metal is present in an electric field $E$, there is an induced charge. If the bound surface charges are not fully compensated, there will be an electric field outside the sample, and therefore an induced charge on the metal. We will assume that the metal indenter is always grounded and calculate the induced charge. For completeness, our calculation considers a full metal sphere. However, because the bound surface charges are largest near the edge of contact, most of the induced surface charge density $\varsigma_m$ on the metal is concentrated near the edge of the contact. Thus, a metal sphere still approximates an asperity contact. This charge may be transferred from the semiconductor or may be simply a redistribution of charge on the indenter, and therefore may or may not be involved in charge transfer; we will discuss both cases later.

A charge $q$ a distance $d$ from the center of a metal sphere of radius $R$ induces a charge $q' = -R/d$ [63]. Therefore, making the same assumption as the contact mechanics, i.e. that $R$ is large compared to the contact radius $a$, the total charge induced $\Delta Q_m$ by an uncompensated surface charge $\varsigma$ is

$$\Delta Q_m = 2\pi \int_a^\infty -\frac{R}{\sqrt{R^2+\rho^2}}\varsigma(\rho)\, \rho d\rho \tag{11}$$

Note that if the interface charge was hypothetically not compensated, because we assume $R \gg a$, the portion inside the contact area is

$$\Delta Q_m = 2\pi \int_0^a -\varsigma(\rho)\, \rho d\rho \tag{12}$$

That is, $\Delta Q_m$ due to the interface charge is the same as compensating the interface charge. Therefore, we assume that the interface is always compensated and ignore the interface in the consideration of the metal charge.

For cylindrical and conical indenters, the calculation of $\Delta Q_m$ is not as simple. Even with simplifying assumptions such as $R \gg a$, the problem is not analytically solvable. Thus, we assume the sphere solution for those indenters. This is likely an acceptable approximation, as most of the integrated $\varsigma$ is concentrated near the edge of contact (i.e., $\rho \approx a$) and is therefore very close to the indenter (see Appendix 4 for detail); in that case, the macroscopic shape of the indenter is of little consequence to how much charge on the metal is induced by the uncompensated surface charge. For example, for the sphere case, $\Delta Q_m$ differs from $\Delta Q_{if}$ by only ~3%, implying the details of the shape have only a small effect on $\Delta Q_m$.

For the 2D case of the roller, instead consider that a linear charge density $\lambda$ induces a linear charge density $\lambda' = -\lambda$ on the roller [63]. Therefore, in these cases,

$$\Delta Q_m = \int_a^\infty -\varsigma(\rho)d\rho \tag{13}$$

As above, we assume this solution approximately holds for the punch.

## 3 Results
### 3.1 Charge Transfer Cases and Sphere Indenter Results

First, we will discuss some bounding cases for surface compensation and charge; real contacts are expected to be between these bounds. As we have mentioned above, the cases depend first on whether the surface charge is entirely compensated; if it is not, then the metal charge is also relevant. Then, there are further cases depending on whether the free interface, surface, or metal charges are involved in charge transfer. Note that, to approximate asperity contact, we assume that the metal is grounded (i.e., it is the end of an asperity connected to a much larger body). Therefore, free charges may be sourced from the metal or the semiconductor bulk. A variety of possible cases are schematically shown in Fig. 3 and discussed below.

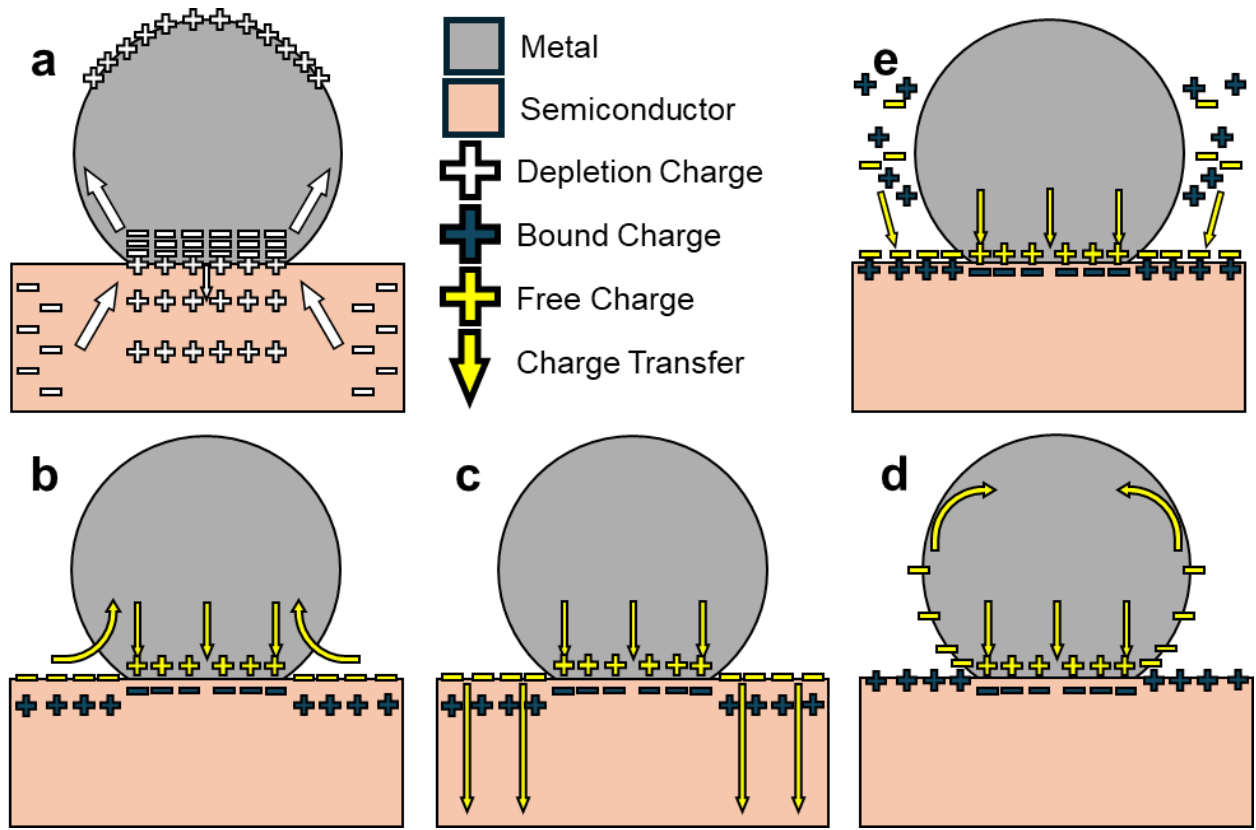

Fig. 3. Schematic diagrams of various cases of charge transfer. Depletion charges are shown in white, bound charges caused by flexoelectricity in dark blue, and free charge in yellow. The arrows show the direction of transfer of positive charges (i.e., current). The sign of bound charges are accurate for the Pt$_{0.8}$Ir$_{0.2}$/Nb:SrTiO$_3$ case, but it will differ for different contacts. (a) Depletion charge transfer always occurs, more details in Fig. 1. (b-e) At the interface, free charges from the metal compensate the bound interface charges and may remain trapped on the semiconductor, causing net charge transfer, here toward the semiconductor. The surface bound charge is (b) compensated by free charge from the metal conducting along the semiconductor surface, (c) charge from the semiconductor bulk, (d) uncompensated, or (e) compensated by atmospheric ions.

We will assume the depletion term always contributes completely to charge transfer, as in Fig. 3a. That is, $+\Delta Q_d$ is included in the total charge transfer in all cases.

At the interface, depending on the materials involved, due to the depletion field $E_d$, the compensating free charges will be sourced from the metal when $\Delta Q_{if} > 0$ as in Fig. 3, or possibly from the semiconductor when $\Delta Q_{if} < 0$. After contact, they may also remain on either the metal or semiconductor. The latter occurs if the charge is trapped in interface states [64], while the former occurs when the Schottky barrier is slightly below the semiconductor surface [25, 65], making transfer back into the bulk semiconductor unlikely. When the free charges are sourced from the semiconductor and remain on the metal, $+\Delta Q_{if}$ is contributed to the total charge transfer, while in the opposite case (Fig. 3b-e), $-\Delta Q_{if}$ is contributed. For brevity, cases where the charge is sourced from and remains on the same body are not shown.

At the (free) surface, the bound charges may or may not be compensated by free charges. If they are, similar to the interface, this charge may be sourced from the metal and conducted along the semiconductor surface (Fig. 3b) or from the bulk of the semiconductor (Fig. 3c). Unlike the interface, the source will not strongly depend on the sign of the bound charge; instead, it will depend on the bulk conductivity of the semiconductor compared to the surface conductivity and the ability of charge to tunnel from the metal to the semiconductor. After contact, the charge will likely be trapped in surface states or conducted into the semiconductor bulk. Thus, when the charge is sourced from the metal, $-\Delta Q_{sf}$ is added to the total transfer (as depicted in Fig. 3b, note that $\Delta Q_{sf} < 0$ here). For case Fig. 3c the free charges on the semiconductor surface will probably just move back into the bulk.

We have considered multiple cases for the surface and interface charge compensation because it is unknown which occurs in Pt$_{0.8}$Ir$_{0.2}$/Nb:SrTiO$_3$ contacts, and more importantly, because which cases correspond to reality will depend on the contacting materials. For example, Ohmic contacts may compensate charges differently than Schottky contacts. Also, we consider the extreme case of complete (or no) charge transfer; real systems may involve partial transfer.

An alternative needs to be considered at the surface. The bound charge (equivalently, surface polarization) may also be compensated by other sources, such as atmospheric ion adsorption (Fig. 3e) or changes in surface reconstruction that accommodate the surface polarization. In these cases, we will assume there is no contribution to charge transfer at the surface, though it is possible that when ions are adsorbed, some charge could be transferred to the semiconductor indirectly by chemical reactions at the surface.

Finally, the bound charge at the surface may be left uncompensated, in which case a charge is induced on the metal (Fig. 3d). Depending on the details of the depletion region, possible tunneling across the

Schottky barrier, and the sign of $\Delta Q_m$, this may result in no charge transfer if charge on metal is simply redistributed (as in Fig. 3d) or charge transfer from the semiconductor to the metal of $+\Delta Q_m$.

For specific numbers, we will consider a contact between a rigid spherical $Pt_{0.8}Ir_{0.2}$ indenter with radius 60 nm, load 4 µN and 0.7 wt% Nb:SrTiO$_3$; this contact is comparable to a single asperity contact or AFM tip contact [25]. Appendix 5 contains a complete table of model parameters. In this case, the values for the depletion and compensating charges are:

- Depletion charge: $\Delta Q_d = 189$ e
- Interface flexoelectric charge: $\Delta Q_{if} = 693$ e
- Surface flexoelectric charge: $\Delta Q_{sf} = -648$ e,

where e is the elementary charge. Therefore, the total charge transfer for each of the cases in Fig. 3 is

$$\text{Fig. 3b:} \quad \Delta Q_{tot}^b = \Delta Q_d - \Delta Q_{if} - \Delta Q_{sf} = 144 \text{ e} \tag{14a}$$

$$\text{Figs. 3c-e:} \quad \Delta Q_{tot}^c = \Delta Q_{tot}^d = \Delta Q_{tot}^e = \Delta Q_d - \Delta Q_{if} = -504 \text{ e} \tag{14b}$$

(The induced metal charge is $\Delta Q_m = -627$ e, but does not play a role in charge transfer.) These numbers indicate that if conduction across the surface is fast (Fig. 3b), the flexoelectric contributions approximately cancel and the dominant term is the depletion charge. This would be relevant for contacts involving metals and/or highly doped semiconductors. In other cases, the transfer is a combination of the contact potential and flexoelectric contributions. Please note that the sign of the flexoelectric coefficient can vary; for SrTiO$_3$ (001) it is positive whereas for SrTiO$_3$ (110) it is negative [59].

### 3.2 Indenter Shapes and Scaling

The calculation is repeated for multiple indenter shapes with an indentation force $F$; in addition to the sphere, we consider a cylinder of radius $R$, a cone with half-angle $\xi$, an infinitely long roller of radius $R$, and an infinitely long square punch with side length $2R$, as seen in Fig. 4. Collectively, the first three are referred to as 3D cases and the latter two as 2D cases. In the 3D cases, $\Delta Q$ has units of charge, whereas in the 2D cases, it has units of charge per unit length.

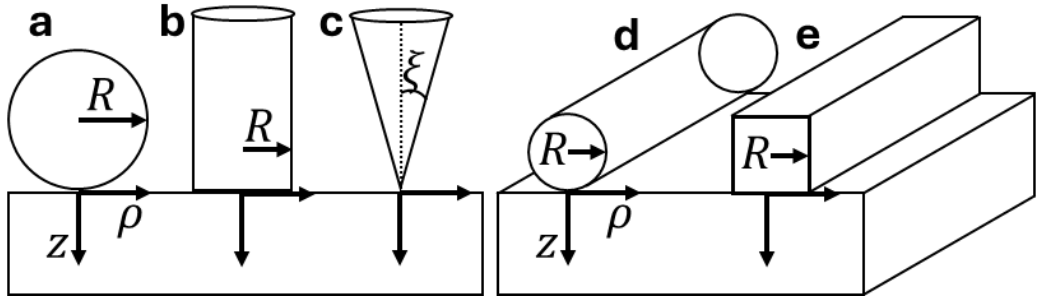

Fig. 4. Coordinate and geometric definitions for indenter shapes: (a) sphere, (b) cylinder, (c) cone, (d) roller, and (e) punch.

As shown in Section 2.1, the depletion charge scales as the contact area $A$ (or contact area per length for the 2D cases), with $\Delta Q_d = -0.81A$ e/nm$^2$ and $A = \pi a^2$ for the 3D cases and $A = 2a$ for the 2D cases. The bound surface charge $\varsigma = P \cdot \hat{n}$ clearly scales with the polarization, so $\Delta Q_{if}$, which is $\varsigma$ integrated over the contact, scales as the polarization times the contact area, i.e., $\Delta Q_{if} \propto AP$. As we noted in section 2.2, the total bound surface charge $\varsigma$ must be opposite the total bound bulk charge $\varrho$, so $\Delta Q_{sf}$, which is $\varsigma$ integrated outside the contact, scales in the same way. That is, $\Delta Q_{if} \propto \Delta Q_{sf} \propto AP \propto Ap_m\mu/aY$ [58]. As discussed in Section 2.3.4 and calculated in Section 3.1, $\Delta Q_m \approx \Delta Q_{sf}$ and scales approximately as $\Delta Q_{sf}$ does.

Normalized results for each shape are given in Table 1, along with the detailed parameters for the calculations. For normalization, we use the value $\mu = \mu_{iiii} = -380$ nC/m. The $\mu$ in the scaling terms refers to even scaling of all the flexoelectric tensor components; relative changes between, e.g., $\mu_{iiii}$ and $\mu_{iijj}$, would of course change $P$. Also, note that in the cases of the cylinder, cone, and punch, contact mechanics solutions that incorporate gradient elasticity [66, 67] were used; else, in particular locations such as the edge of contact for the cylinder, the strain gradients, and therefore the polarization and bound charges, are extremely large due to the sharp corners of the indenters. The gradient elasticity scale for SrTiO$_3$ has been calculated as $\ell = 4.04$ Å [68].

|  | Geometry | F | $A = \dfrac{\Delta Q_d}{0.81 \text{ e/nm}^2}$ | $\dfrac{\Delta Q_{if} aY}{Ap_m\mu} \propto \dfrac{\Delta Q_{if}}{AP}$ | $\dfrac{\Delta Q_{sf} aY}{Ap_m\mu} \propto \dfrac{\Delta Q_{sf}}{AP}$ |
|---|---|---|---|---|---|
| Sphere | $R = 60$ nm | 4 µN | 232 nm² | -0.168 | 0.157 |
| Cylinder | $R = 10$ nm | 4 µN | 314 nm² | -0.158 | 0.145 |
| Cone | $\xi = 45°$ | 4 µN | 28 nm² | -0.121 | 0.111 |
| Roller | $R = 60$ nm | 300 N/m | 18 nm | -0.244 | 0.064 |
| Punch | $R = 10$ nm | 400 N/m | 20 nm | -0.103 | 0.144 |

Table 1. The geometry and force $F$ used for the calculations are given, along with the contact area $A$, and consequently $\Delta Q_d$, as well as the unitless results $\Delta Q_{if} aY/Ap_m\mu$ and $\Delta Q_{sf} aY/Ap_m\mu$, where $a$ is the contact radius, $Y$ the semiconductor modulus, $p_m$ the mean contact pressure, and $\mu = -380$ nC/m the flexoelectric coefficient.

The scaling of $\Delta Q_d$ relative to $\Delta Q_{if}$ and $\Delta Q_{sf}$ implies that, when both are involved in charge transfer, there is a contact regime where the depletion term will dominate and another where the flexoelectric terms will dominate. The relative scaling

$$\frac{\Delta Q_d}{\Delta Q_{sf}} \propto \frac{\Delta Q_d}{\Delta Q_{if}} \propto \frac{A}{AP} \propto \frac{1}{P} \propto \frac{aY}{p_m\mu} \tag{15}$$

indicates that, simply, when the polarization is large, the flexoelectric terms dominate. To get a sense for what is "large" for the Pt$_{0.8}$Ir$_{0.2}$/Nb:SrTiO$_3$ case, we note that the magnitudes of $\Delta Q_{sf}/AP$ and $\Delta Q_{if}/AP$ are about 0.1, while $|\Delta Q_d/A| = 0.81$ e/nm². Therefore, the flexoelectric terms begin to dominate by a factor of 10 when $|P| > 8.1$ e/nm² $= 1.3$ C/m². For better context, using $\mu = -380$ nC/m, this corresponds to a strain gradient of 3.42 µm$^{-1}$, which is large but not impossible at asperity contacts.

Until now, to preserve generality, we have discussed the scaling of the charges in terms of the contact radius $a$, contact area $A$, mean pressure $p_m$ and polarization $P$. The force $F$, modulus $Y$ and radius $R$ or half-angle $\xi$ are much more intuitive quantities. For the shapes considered in this section, Table 2 provides conversions between these terms. Note that the indenter shape vastly impacts how the contact area and polarization, and therefore the charge transfer, scale with the indentation force and size. This is important in practice, as we will detail later.

| Indenter Shape | $a$ | $\Delta Q_d \propto A$ | $p_m = F/A$ | $P \propto p_m \mu / aY \propto$ | $\Delta Q_{if} \propto \Delta Q_{sf}$ $\propto AP \propto$ |
|---|---|---|---|---|---|
| Sphere | $\left(\frac{3}{4}(1-\nu^2)\frac{RF}{Y}\right)^{\frac{1}{3}}$ | $\pi a^2$ | $\left(\frac{16FY^2}{9\pi^3(1-\nu^2)^2 R^2}\right)^{\frac{1}{3}}$ | $\frac{\mu}{R}$ | $\mu\left(\frac{F^2}{RY^2}\right)^{\frac{1}{3}}$ |
| Cylinder | $R$ | $\pi a^2$ | $\frac{F}{\pi R^2}$ | $\frac{\mu F}{R^3 Y}$ | $\frac{\mu F}{RY}$ |
| Cone | $\left(2(1-\nu^2)\frac{F \tan \xi}{Y}\right)^{\frac{1}{2}}$ | $\pi a^2$ | $\frac{Y}{2\pi(1-\nu^2)\tan \xi}$ | $\mu\left(\frac{Y}{F\tan^3 \xi}\right)^{\frac{1}{2}}$ | $\mu\left(\frac{F}{Y\tan \xi}\right)^{\frac{1}{2}}$ |
| Roller | $\left(\frac{4}{\pi}(1-\nu^2)\frac{FR}{Y}\right)^{\frac{1}{2}}$ | $2a$ | $\left(\frac{\pi YF}{16(1-\nu^2)R}\right)^{\frac{1}{2}}$ | $\frac{\mu}{R}$ | $\mu\left(\frac{F}{RY}\right)^{\frac{1}{2}}$ |
| Punch | $R$ | $2a$ | $\frac{F}{2R}$ | $\frac{\mu F}{R^2 Y}$ | $\frac{\mu F}{RY}$ |

Table 2. Scaling of the contact radius $a$, contact area $A$, mean contact pressure $p_m$, polarization $P$, depletion charge $\Delta Q_d$, and flexoelectric interface and surface charges $\Delta Q_{if}$ and $\Delta Q_{sf}$ with respect to the contact force $F$, indenter radius $R$ or half-angle $\xi$, Young's modulus $Y$, Poisson's ratio $\nu$, and flexoelectric coefficient $\mu$. Note that for the 2-D roller and punch cases, $F$ has units of force per length, $A$ length, and the charges $\Delta Q$ charge per length.

To clarify the difference in scaling between the indenter shapes, we plot $\log_{10}[2|\Delta Q_d|/(|\Delta Q_{if}| + |\Delta Q_{sf}|)]$ over regions of asperity geometry ($R$ or $\alpha$) and normalized force $F/Y$ for the various indenter shapes in Fig. 5. This is a measure of the relative influence of depletion and flexoelectric terms which is, e.g., 0 (the solid lines in Fig. 5) when the two terms have the same magnitude, and 2 when the flexoelectric terms are roughly 1% of the depletion term. Light colors represent regions where the depletion charge is dominant in all cases. In dark regions, the flexoelectric terms dominate, except in the cases where $\Delta Q_i$ and $\Delta Q_s$ largely cancel each other, such as in the sphere case corresponding to Fig. 3b, where the flexoelectric contribution to the total charge transfer is $(\Delta Q_{if} + \Delta Q_{sf})/AP = -0.168 + 0.157 = -0.011$. For this example, the flexoelectric terms will only be larger when $\log_{10} 2|\Delta Q_d|/(|\Delta Q_{if}| + |\Delta Q_{sf}|) \lesssim -1$, i.e., only where Fig. 5 is nearly black.

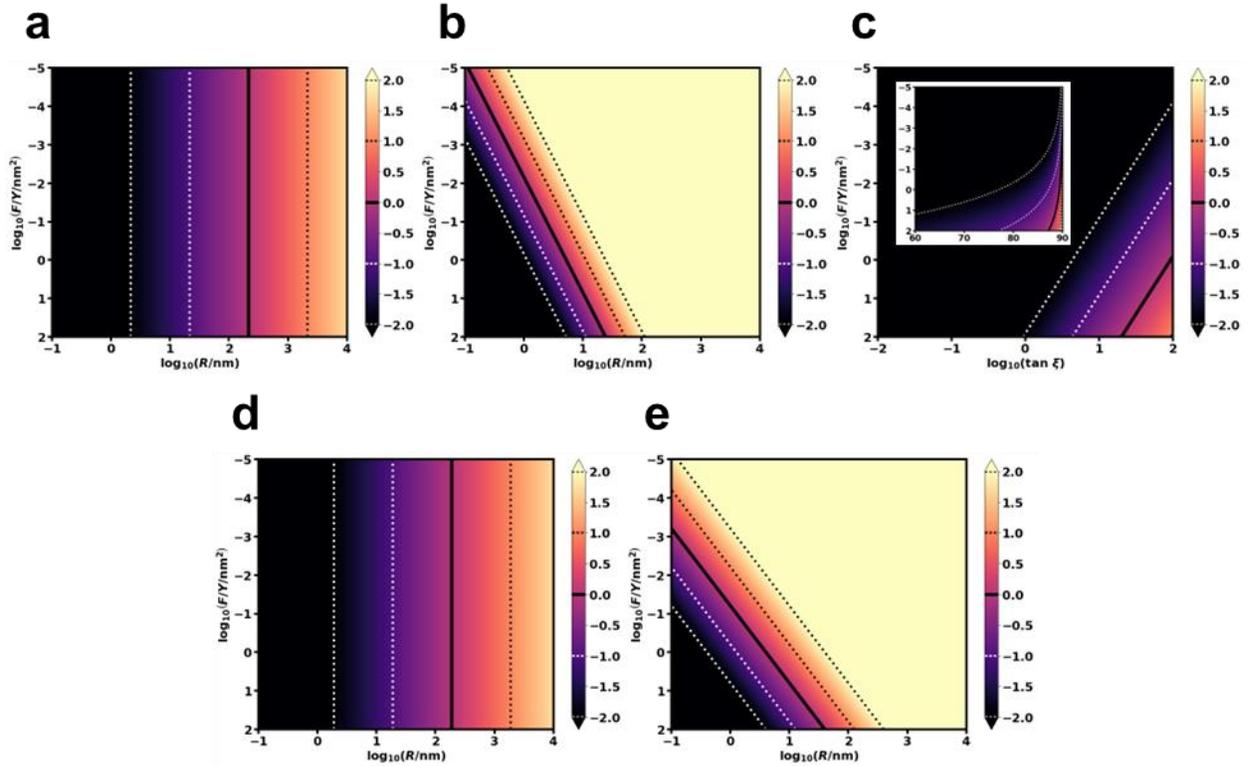

Fig. 5. Plots of $\log_{10} 2|\Delta Q_d|/(|\Delta Q_{if}| + |\Delta Q_{sf}|)$ for (a) sphere, (b) cylinder, (c) cone, (d) roller, and (e) punch indenters with various indenter geometries and normalized indentation forces $F/Y$. The inset in (c) has $\xi$ in degrees on the x axis and the same normalized indentation force on the y-axis.

## 4 Discussion

We have attempted to provide a comprehensive analysis of contact charge transfer. Because there are multiple pathways there is no simple "one equation fits all"; everything depends heavily upon the properties of the materials. Depending upon the exact nature of the contact, either the contact potential or the flexoelectric terms can dominate, or triboelectric charging is a combination of the two. It is known that often experimental tribocharge measurements are hard to reproduce [14, 20]. Our analysis indicates that this is fundamental to the science, and only with very highly controlled systems will there be a high level of reproducibility.

We have focused on one specific material pair to build out a detailed theory, but the ideas apply for other contacts. Those between a metal and a crystalline semiconductor will follow closely. We considered Schottky contacts and Ohmic contacts have some differences, for instance the lack of a barrier. However, there will still be charge transfer to align the bands (analogous to $\Delta Q_d$) so the process can also be described by analogy to a capacitor. The electric current behavior of Ohmic and Schottky contacts are different, so the extent that the band lineup does contribute to the net transfer after contact may differ. For contacts between a metal and a perfect, zero-doped insulator, the theory would be different. However, real insulators involved in triboelectric contacts have defects such as dopants or non-stoichiometry and their band structures are not fundamentally different from semiconductors, except that the band gap may be larger. Similarly, polymers have band structures that are not principally different from crystalline semiconductors or defective insulators. Because flexoelectricity is present in all non-metals, it is also relevant for any of these material classes.

In Section 3.2, we analyzed the effect of indenter shape on the charge transfer. The indenter shape significantly changes how the charge transfer scales with the contact details. We have analyzed simple geometries; they are bounds for real asperity shapes [69] and are directly relevant for artificially constructed surfaces that are used to improve triboelectric nanogenerator performance [70].

It is relevant to connect our analysis to experimental data, which we will do in the next paragraphs. At the same time we will point towards some of the (many) issues that remain unresolved, which are topics for future work.

Our results for the magnitude of charge transfer, see Eq. (14), agree roughly with a previous empirical model based on the condenser theory [71]. We consider a contact between a 500 nm metal particle and a large (flat) 600 µm silicate chondrule precursor particle, which are precursors to millimeter-sized glassy beads in meteorites and are commonly comprised of silicates such as olivine or pyroxene, as described in ref. [71]. The empirical model of that work predicted a charge transfer of 2700 e, approximately applying our model to this contact gives $\Delta Q_d = 6.7$ e, $\Delta Q_{if} = -806$ e, and $\Delta Q_{sf} = 741$ e. More detail of this calculation is given in Appendix 6. The magnitudes of these predictions are similar, showing that our model is in rough agreement with previous models. We will discuss shortly why a factor of three can be considered similar in the context of quantitative triboelectric modeling.

The condenser model of charge transfer predicts charging scales with the contact area. Our scaling results agree when the depletion potential and the associated charge $\Delta Q_d$ dominate. However, when the flexoelectric terms are relevant (for spherical indenters, when the radius is small), the charge transfer scales with the polarization multiplied by the contact area. In experiments involving particle impacts [2, 72, 73], the charge transferred appears to scale with the velocity $v$ as $c_0 + c_1 v$, where $c_i$ are constants. The Hertz solution for contact area during impacts scales as $v^{4/5}$, but for plastic collisions the proportionality increases to $v$ [2], indicating that the charge transfer scales with contact area in these cases. Experiments that directly compare the charge transfer to contact area also find that charge transfer scales with the contact area [2, 74]. This agrees with our model in the limit that $\Delta Q_d$ is dominate. As discussed in Section 3.2, for spheres this corresponds to the cases with relatively large particle radii.

Scaling due to the flexoelectric terms has not typically been observed in experiments because the diameter of the impacting particles was in the range 100 µm – 10 mm [2, 72-78], while experiments with AFM techniques used 5 µm spheres [79]. According to Fig. 5a, assuming flexoelectric coefficients similar to SrTiO$_3$, the depletion transfer is dominant even in the smallest case, and especially so for the typical particle size range. The only experiments we are aware of that directly measure triboelectricity in nanoparticles involve only metal/metal contacts [80, 81]. Therefore, while the flexoelectric terms are significant in the contact studied here, a single asperity, they are insignificant in most currently published particle impact experiments. Flexoelectric terms remain significant in many cases involving contacts of small asperities, though separating the impact of depletion and flexoelectric terms quantitatively is much more challenging for macroscopic contacts which involve many asperity contacts. Various experiments dating back to the early 20$^{th}$ century show phenomena that can be explained by flexoelectric terms, and not simply by the depletion term. For example, Jamieson in 1910 and Shaw in

1917 found that the magnitude and sign of charge transfer between changed depending on the macroscopic curvature or pre-strain of the contacting materials [12, 13]. This has continued to be confirmed in various experiments, such as the sign and magnitude of charge transfer during impacts between PTFE spheres and latex sheets depending on the pre-strain of the sheets [82, 83].

Many experiments of single particle impacts [2, 72-79] show a charge transfer per contact area of $\sim 10^{-2} - 10^{-4}$ e/nm$^2$ for single contacts, mostly between polymer powder particles and metal targets. Note that the Hertz solution for contact area during impacts [53] was used to estimate the contact area where it was not reported in the experiments.

Unfortunately, there is very little data on single metal/semiconductor contacts. One experiment [84] measured many glass beads impacting an aluminum plate. Applying our model to this problem (with many rough assumptions), the predicted charge transfers are $\Delta Q_d = 3.78 \cdot 10^6$ e, $\Delta Q_{if} = -1.50 \cdot 10^5$ e, and $\Delta Q_{sf} = 1.38 \cdot 10^5$ e per contact. The charge transfer observed in this experiment was, on average, about $10^6$ e per contact, which demonstrates approximate quantitative agreement with experiments. Further details are found in Appendix 6.

One possible reason for the quantitative discrepancy is that, in many cases, single contacts are not the end of charge transfer. Studies with repeated contacts between particles [82, 83, 85, 86], in contact-mode triboelectric nanogenerators [87, 88], or in other cases [89, 90] show charge transfer saturating only after many contacts; typically, the charge transfer in the first contact appears to be ~10% or less of the saturated value. The values presented here do not consider incomplete transfer. The assumptions of complete charge transfer and that the dynamic mechanism described in Fig. 1 was dominated entirely by the charge transfer depicted in Fig. 1d both explain why the calculated value differs from experimental results. For example, interface or surface trap states may hinder complete charge transfer. The details of dynamic depletion region formation during contact and the resulting charge transfer are complex and not well-studied, and further work in this area would provide further insight into charge transfer mechanisms and better quantitative models.

Further quantitative inaccuracy is introduced by the simplification of the contact mechanics. In the model presented here, we have considered Hertzian contact with a rigid indenter. This has the benefit of maintaining simplicity and consistent scaling across a range of length scales but introduces some errors. Considering non-rigid indenters, adhesion between the contacting bodies [91], gradient elasticity [66, 67], surface stress [92], and contact solutions that consider coupling between flexoelectricity and stress [93, 94] each or all may be important, depending on the contact. These corrections result in errors of about a factor of 2 or less. That is, these corrections are quantitatively significant, but relatively small given the current state of triboelectric models. Appendix 7 includes more detail and some calculation results involving these corrections. Though coupling between mechanical and electrical portions of the contact problem is important, care must be taken to avoid double counting influences of certain phenomena. For example, the charge due to the formation of a depletion zone at a Schottky contact can be directly counted, as in Section 2.1, or considered as an apparent piezoelectric polarization due to the gradient of the band-bending (local electric field), as in ref. [36]. Apparent piezoelectricity can also appear to due converse flexoelectricity [95]. Including both the apparent piezoelectricity due to band bending and the depletion term directly, or both that due to converse flexoelectricity and the flexoelectric effect directly, would double count the effects of the depletion.

Another factor that can vary between different materials is the depletion charge $\Delta Q_d$, which is affected by the doping level $N$, the permittivity $\epsilon$, and the potential difference between the contacting materials $V_d$. Recall from Section 2.1 that $\Delta Q_d = \epsilon_0 \gamma \sqrt{\delta^2 - 1}/\delta$ where $\delta = \beta V_d/\alpha + 1 = 1 + eNV_d/\epsilon_0 \alpha \gamma$ and $\epsilon = \epsilon_0 \gamma/\sqrt{\alpha^2 + E^2}$. When $eNV_d/\epsilon_0 \alpha \gamma \ll 1$, $\Delta Q_d \approx \epsilon_0 \gamma \sqrt{2eNV_d/\epsilon_0 \alpha \gamma}$, and thus $\Delta Q_d$ is small and increases with the square root of the doping level, permittivity, and work function difference. On the other hand, when $eNV_d/\epsilon_0 \alpha \gamma \gg 1$, $\Delta Q_d \approx \epsilon_0 \gamma$, and thus $\Delta Q_d$ increases only with the permittivity. As an example, changing $N$ from 2.24 × $10^{26}$ m$^{-3}$ to 2.24 × $10^{24}$ m$^{-3}$, i.e., 0.7 wt% Nb:SrTiO$_3$ to nearly undoped SrTiO$_3$, we find the depletion charge changes from $\Delta Q_d = 0.81$ e/nm$^2$ to 0.34 e/nm$^2$. Similarly, reducing the permittivity from $365\epsilon_0$ for Nb:SrTiO$_3$ to $5\epsilon_0$, that of glass [96], results in a drastic decrease to $\Delta Q_d = 0.011$ e/nm$^2$. Properly determining the band structures of contacting materials, in addition to precisely determining the electronic properties, such as the doping level and permittivity of semiconductors involved in the contact, is critical for quantitative predictions.

There has been recent work supporting that doping level changes can significantly change triboelectric charge transfer. For example, the doping level of semiconductors has been shown to affect charge transfer when they contact metals [97-99], other semiconductors [97] including polymers [100], and even liquids [101]. This is typically attributed to doping changing the built-in field or equivalently the depletion region formed by the contacts. Closely related to this work is other recent work that shows doping can enhance effective flexoelectric coefficients in, e.g., three-point bending measurements [102-104]. This is also attributed to the changing depletion region and not an effect on intrinsic flexoelectric properties. These results are deeply connected to each other and the analysis of the depletion region in Section 2.1, but further work is needed to determine how exactly they are linked and whether additional theory is necessary.

The assertion in Section 3.1 that the interface charge may be trapped in the semiconductor interface states or surface states is reasonable only if those states are available in a density comparable to the density of $\Delta Q_d$, $\Delta Q_{if}$, and $\Delta Q_{sf}$. The interface state density for Pt/Nb:SrTiO$_3$ has been measured to be 0.35 eV$^{-1}$ nm$^{-2}$ [64]. For the values reported in Eq. (14), the contact radius is 8.6 nm, so the total number of interface states is 81 eV$^{-1}$. This value is comparable to $\Delta Q_d$ and $\Delta Q_{if}$, and therefore assuming complete charge transfer does not require an unreasonable interface state density. Similarly, the surface state density for SrTiO$_3$ is roughly 0.1-0.5 nm$^{-2}$ [105-108], and the surface area over which the surface charge is distributed is roughly 25 times the contact area, so there are ~600-3000 surface states available, which is comparable to or larger than $\Delta Q_s$.

The impact of humidity on triboelectricity may support the importance of the surface charge $\Delta Q_s$ in charge transfer. In many cases, higher humidity leads to decreased charge transfer [109, 110]. At higher humidities, more ions can be adsorbed onto the surface [111], thereby moving from, for example, case b to case e in Fig. 3. Correspondingly, we calculate that the charge transfer will decrease from $144$ e to $-504$ e for the extreme cases, see Eq. (14). That is, the drop in charge transfer at high humidities corresponds with our results as a higher fraction of the surface is compensated by atmospheric ions. Humidity interacting with flexoelectric terms may explain some seemingly anomalous results in triboelectric experiments, such as the reversal of charging sign with changing humidity [112], charging increasing with humidity [113], or the effect of humidity being stronger for smaller particles [114].

# 5 Conclusion

We have developed a model that, without any fitting parameters, quantitatively calculates the charge transfer that occurs during a metal-semiconductor contact. Charge transfer is a combined result of the contact potential between materials in addition to compensation of flexoelectric bound charges that form at contacts. The model is flexible for different material and environmental conditions, and we have discussed which regions of contact may be involved in charge transfer and when contact potentials or flexoelectricity are the dominant driving force. Our model explains the scaling of charge observed with particle size and force or impact velocity and fits with a drop in charge transfer at high humidity.

## Acknowledgements

KPO was supported by the Northwestern University McCormick School of Engineering.

## Compliance with ethical standards

### Declaration of competing interest

The authors have no competing interests to declare that are relevant to the content of this article.

### Ethical approval

This study does not contain any studies with human or animal subjects performed by any of the authors.

## Appendix 1. Depletion Charge from 3D Potential

In a full 3D solution, we must realize that the depletion potential, and not necessarily the space charge, should be dependent on the distance from the interface. Let us take the potential to be $\psi(z)$ for $0 < \rho < a$ and $\psi\left(\sqrt{z^2 + (\rho - a)^2}\right)$ for $\rho > a$, so that it is the 1D solution where the coordinate is distance from the interface. For simplicity, we define $d$ as the distance from the interface, so this is simply $\psi(d)$. Then, we find that $\left|2\pi \int_0^\infty \int_0^\infty q_{3D}^s(\rho, z) \, \rho d\rho \, dz\right| = 1450$ e when the indenter has radius 60 nm and the force is 4 μN, where $q_{3D}^s(\rho, z) = \epsilon_{STO} \nabla^2 \left(\psi(d) + \psi^{im}(d)\right)$, where $\psi^{im}$ is the image potential, which is $\psi^{im}(\rho, z) = \psi(\rho, -z)$. This image potential arises by assuming, in the context of image terms only, the metal indenter can be approximated by a metal half-space above the semiconductor. This is approximately justified because most of the charge is directly below the metal indenter.

The value of 1450 e found here is comparable to $|Q_d| = \left|2\pi \int_0^\infty \int_0^\infty q^s(d) \, \rho d\rho \, dz\right| = 1213$ e, i.e., when we the charge distribution is the 1-D solution with only the distance from the interface mattering. Note that $q^s(d) \napprox q_{3D}^s(d)$; only their integrals are similar, meaning that approximating the 3D space charge as the 1D space charge depending on the distance is acceptable only if one is concerned with the total space charge.

Note that from the line integral of Eq. (5) and Eq. (6), the total charge change in the semiconductor is also the effective surface bound charge of the depletion potential. Therefore, these terms agreeing is expected.

## Appendix 2. Depletion Charge for a Strained Material

Here, will estimate the coupling between strain and the depletion region in the continuum limit. We assume the 1-D solution as in Section 2.1 but consider a material which is strained by $\varepsilon$. Then, if $W_0$ is the unstrained depletion width as defined in Eq. (1c), the depletion width for the strained system is $W = (1 + \varepsilon)W_0$. The rest of the analysis remains the same, and the depletion charge is $|\Delta Q_d| = \epsilon_0 \gamma \tanh(\beta W)$. Substituting $W = (1 + \varepsilon)W_0$ gives $|\Delta Q_d| = \epsilon_0 \gamma \tanh(\beta(1 + \varepsilon)W_0)$. Taylor expanding around $\varepsilon \approx 0$, i.e., assuming small strains, we have,

$$|\Delta Q_d| \approx \epsilon_0 \gamma [\tanh(\beta W_0) + \beta W_0 \varepsilon \operatorname{sech}^2(\beta W_0) + \mathcal{O}(\varepsilon^2)]$$

where we have assumed small strains and thus ignore terms of order $\varepsilon^2$ and higher. To first order, then, the effect of strain is to change the depletion charge by a factor of

$$r = \frac{\tanh(\beta W_0) + \beta W_0 \varepsilon \operatorname{sech}^2(\beta W_0)}{\tanh(\beta W_0)}$$

where $r = 1$ corresponds to no change. Now, note that

$$\beta W_0 = \cosh^{-1}\left(\frac{\beta}{\alpha}V_d + 1\right) = \cosh^{-1}(\Lambda)$$

and recall $\beta = eN/\epsilon_0\gamma$. Thus, when the doping level is very high, $N$ is large, and consequently $\beta$ and $\Lambda$ is large. In the high doping limit, there is no effect of strain, since

$$\lim_{\Lambda \to \infty} r = \lim_{\Lambda \to \infty} \frac{\tanh(\cosh^{-1}(\Lambda)) + \varepsilon \cosh^{-1}(\Lambda) \operatorname{sech}^2(\cosh^{-1}(\Lambda))}{\tanh(\cosh^{-1}(\Lambda))} = 1$$

Meanwhile, in the zero-doping limit, $\beta = 0$ and $\Lambda = 1$. In this case, the depletion charge is linear in strain, since

$$\lim_{\Lambda \to 1} r = \lim_{\Lambda \to 1} \frac{\tanh(\cosh^{-1}(\Lambda)) + \varepsilon \cosh^{-1}(\Lambda) \operatorname{sech}^2(\cosh^{-1}(\Lambda))}{\tanh(\cosh^{-1}(\Lambda))} = 1 + \varepsilon$$

Recall $|\Delta Q_d| = r|\Delta Q_{d0}|$ where $|\Delta Q_{d0}|$ is the unstrained depletion charge. For the highly-doped Nb:SrTiO$_3$ in the main text, $\beta V_d/\alpha = \Lambda - 1 = 10$, so the strained depletion charge would be $|\Delta Q_d| = (1 + 0.025\varepsilon)|\Delta Q_{d0}|$. That is, the strain has very little effect on the depletion charge in the present case. Therefore, it is ignored in the main text.

## Appendix 3. Estimate of Complete Flexoelectric Tensor Components

Theoretically calculated flexoelectric coefficients, which we will label $\mu'_{ijkl}$ and which have values $\mu'_{1111} = -36.9$ nC/m, $\mu'_{1122} = -40.2$ nC/m, and $\mu'_{1212} = -1.4$ nC/m for SrTiO$_3$ [115], do not include the so-called "surface flexoelectricity" which is caused by gradients in the mean inner potential due to volumetric strain changes [57]. Experimental flexoelectric coefficients measured in the typical three-point bending experiment, $\mu_{\text{eff}}^{(mnp)}$, do include this term, but are effective flexoelectric coefficients for the surface orientation $(mnp)$ normal to the bending direction; they are related to proper tensor components but cannot be completely decomposed without further assumptions [33, 60]. Here, we will use experimental values for $\mu_{\text{eff}}^{(100)}$ and $\mu_{\text{eff}}^{(110)}$ to estimate the three independent flexoelectric tensor components $\mu_{1111}$, $\mu_{1122}$, and $\mu_{1212}$ for SrTiO$_3$. These $\mu_{ijkl}$ do include the mean inner potential term.

Measurements of SrTiO$_3$ give $\mu_{\text{eff}}^{(100)} = 12.4$ nC/m and $\mu_{\text{eff}}^{(110)} = -8.3$ nC/m [59]. These are related to $\mu_{ijkl}$ by

$$\mu_{\text{eff}}^{(100)} = \frac{s_{1122}}{s_{1111}}\mu_{1111} + \frac{s_{1111} + s_{1122}}{s_{1111}}\mu_{1122} - \phi\frac{s_{1122}}{s_{1111}}\left(\frac{s_{1122}}{s_{1111} + s_{1122}}\mu_{1111} + \mu_{1122}\right)$$

$$\mu_{\text{eff}}^{(110)} = \frac{s_{1111} + s_{1122}}{s_{1111} + s_{1122} + 2s_{1212}}\mu_{1111} + \frac{s_{1111} + 3s_{1122}}{s_{1111} + s_{1122} + 2s_{1212}}\mu_{1122} - \frac{4s_{1122}}{s_{1111} + s_{1122} + 2s_{1212}}\mu_{1212}$$
$$+ \frac{\phi}{2}\mu_a^{(110)}$$

$$\mu_a^{(110)} = -\frac{4s_{1122}\mu_{1122}}{s_{1111} + s_{1122} + 2s_{1212}} + \frac{(s_{1111} + s_{1122} - 2s_{1212})s_{1212}(\mu_{1111} + \mu_{1122} + 2\mu_{1212})}{(s_{1111} + s_{1122})(s_{1111} + s_{1122} + 2s_{1212})}$$
$$- \frac{(s_{1111} + s_{1122} - 2s_{1212})(\mu_{1111} + \mu_{1122} + 2\mu_{1212})}{2(s_{1111} + s_{1122} + 2s_{1212})}$$

where $s_{ijkl}$ are the stiffness tensor components and $\phi$ is a parameter that describes the bending geometry in the experiment, ranging from 0 (pure beam bending) to 1 (pure plate bending). For these experiments, $\phi \approx 0.1$.

This is presently unsolvable; there are more unknown quantities ($\mu_{1111}$, $\mu_{1122}$, and $\mu_{1212}$) than known quantities ($\mu_{\text{eff}}^{(100)}$ and $\mu_{\text{eff}}^{(110)}$). To emphasize this point, a measurement of another surface orientation would not solve this problem, as it would simply be a linear combination of the other two measurements. To proceed, we will assume the solutions take the form of

$$\mu_{1111} = \mu'_{1111} + \mu_1^s$$
$$\mu_{1122} = \mu'_{1122} + \mu_2^s$$
$$\mu_{1212} = \mu'_{1212}$$

where $\mu_i^s$ are effective surface flexoelectric terms. Under this assumption, we find $\mu_1^s = -343$ nC/m and $\mu_2^s = -63$ nC/m.

For a volumetric strain $\varepsilon_{\text{vol}}$, the mean inner potential (MIP) shifts relative to the vacuum energy $E_{\text{vac}}$. In SrTiO$_3$, this shift is

$$\varphi = \frac{\partial(E_{\text{MIP}} - E_{\text{vac}})}{\partial \varepsilon_{\text{vol}}} \approx 22.2 \text{ V}$$

by the Ibers approximation [116]. Thus, the potential due to this term is $V = \varphi\varepsilon_{\text{vol}}$, which can be related to a polarization, and thus included in the flexoelectric polarization, by $\epsilon\varphi\nabla(\varepsilon_{\text{vol}})$, where $\epsilon$ is the dielectric permittivity. Thus, our assumed form for the flexoelectric components is justified by two points:

1. The MIP contributes to polarization via volumetric strain gradients, so $\mu_{1212}$ does not vary from theory, as the shear strain $\varepsilon_{12}$ does not contribute to volumetric strain. Thus, $\mu_{1212} = \mu'_{1212}$.
2. The magnitude of $\mu_i^s$ roughly match the magnitude of $\epsilon\varphi = 71.8$ nC/m. That is, the introduced effective surface flexoelectric coefficients do not vary perversely from established estimates of the surface flexoelectric effect by the MIP shift.

## Appendix 4. Metal Charge Induction Detail

Here, we show that most of the integrated surface charge $\varsigma$ is indeed near the contact, implying that the shape of the indenter has little effect on the calculation of the metal charge.

Figure A1 shows a plot of the contribution of the integral up to $\rho$, that is,

$$\Delta Q_m(\rho < r) = 2\pi \int_a^r -\frac{R}{\sqrt{R^2 + \rho^2}} \varsigma(\rho) \, \rho d\rho$$

To clarify, $\Delta Q_m(\rho < r)$ is the total induced metal charge when considering only the surface charges $\varsigma$ in the range $a < \rho < r$, so that the total induced charge is $\Delta Q_m = \Delta Q_m(\rho < \infty)$.

Figure A1 shows that $\Delta Q_m(\rho < r)$ becomes close to $\Delta Q_m = -627$ e at a distance of $0.5R$ away from the contact. That is, nearly all the contribution to the induced metal charge comes from surface charges inside the region from the contact edge to half the indenting sphere radius away. Because this region is very close to the the indenter, this indicates the shape of the sphere radius is largely unimportant to the calculation of $\Delta Q_m$.

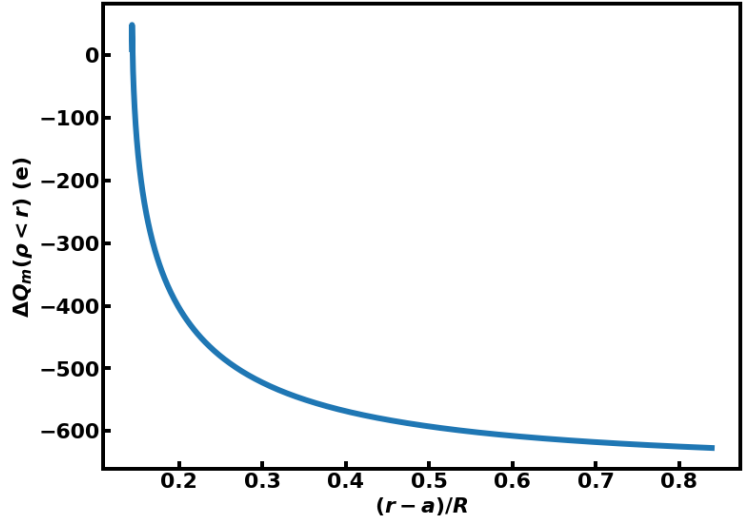

Figure A1. A plot of $\Delta Q_m(\rho < r)$, the cumulative contribution of the surface charge $\varsigma$ in the region up to a distance $r$ outside the edge of contact to the induced metal charge $\Delta Q_m$, against that distance $r$ normalized by the sphere radius $R$. Most of the induced metal charge is caused by surface charges less than $0.5R$ from the contact edge.

## Appendix 5. Table of Model Parameters

| Parameter | Value (units) | Reference |
|---|---|---|
| $F, R$ | 4 (µN), 60 (nm) | - |
| $Y, \nu$ | 270 (Gpa), 0.24 | [54] |
| $\alpha, \gamma$ | 0.04, 14.8 (V/nm) | [42] |
| $\epsilon = \epsilon_0 \gamma / \sqrt{(a^2 + E^2)}$ | | |
| $N$ | 2.24 × $10^{26}$ (m$^{-3}$) | - |
| $\Phi_b$ | 1.4 (V) | [25, 43-46] |
| $\zeta_F$ | 0.065 (V) | Calculated, [47] |
| $\mu_{1111}, \mu_{1122}, \mu_{1212}$ | -380, -103, -1.4 (nC/m) | See Appendix 3 |

Table A1. List of model parameters and values. The indenter force $F$, indenter radius $R$, and doping level $N$ were chosen to match the AFM contact in [25], though these are adjustable.

# Appendix 6. Estimated Calculations with Other Materials

For comparisons to the empirical theory of [71] and the experiments in [84], we have roughly applied the model to contacts between a metal and silicate minerals and between a metal and silica glass, respectively. We have assumed that all parameters remain the same except:

1. Both cases involve impacts between particles. The indenter radius $R$ is set to the radius of smaller of the impacting bodies. For the two cases, these are 500 nm and 275 µm, respectively. We have assumed that the large silicate chondrule precursor is effectively flat, as it is 600 µm compared to the 500 nm metal particle.
2. The maximum indenting force $F$ is determined by Hertzian mechanics [53]. The impact velocities in the two cases are 0.5 m/s and 1 m/s, respectively. This corresponds to forces of 310 nN and 0.21 N, respectively.
3. The appropriate mechanical properties are used. The Young's modulus and Poisson's ratio for silicates and silica glass are about 100 Gpa and 0.27, respectively [117-119]. The authors of [84] give values of 75 GPa and 0.17 for their glass particles.
4. The flexoelectric coefficients for the silicate particles are assumed to be those of $SiO_2$, $\mu_{11}=\mu_{12}=10\mu_{44}=-110$ nC/m [120]. Those for the glass particles are assumed to be $\mu_{11}=\mu_{12}=10\mu_{44}=-25$ nC/m based on our own unpublished preliminary three-point bending flexoelectric measurements.
5. The field-dependent permittivity, $\epsilon = \epsilon_0 \gamma/\sqrt{\alpha^2 + E^2}$, is used, where $\alpha$ is held constant and $\gamma$ is adjusted so that $\epsilon_{\text{silicates}} \approx 8\epsilon_0$ [121] and $\epsilon_{\text{glass}} \approx 5\epsilon_0$ [96].

Note that we have not included a new doping level $N$ or work function difference $V_d$. In the highly-doped case (whether the materials are purposefully doped or are highly defective), $V_d$ has little impact on $\Delta Q_d$. Chondrule precursor particles undergo a variety of processes that affect their stoichiometry [122]. There is little detail given about the glass beads in [84], but commercial glass typically contains various additives. Therefore, the assumption of relatively high doping is reasonable in both cases.

With these modified parameters, we find $\Delta Q_d = 6.7$ e, $\Delta Q_{if} = -806$ e, and $\Delta Q_{sf} = 741$ e as a comparison to [71] and $\Delta Q_d = 3.78 \cdot 10^6$ e, $\Delta Q_{if} = -1.50 \cdot 10^5$ e, and $\Delta Q_{sf} = 1.38 \cdot 10^5$ e as a comparison to [84]. Note that in the first case there is a much larger flexoelectric contribution as the particle radius is much smaller (see Section 3.2); in the second case the contact potential term is dominant.

From equation 14 in [71], the previous empirical model predicts, for a particle of radius 500 nm and a potential difference of ~2 V, a charge transfer of 2700 e is predicted.

For the experiments in [84], a current of about $I = 0.5$ nA is measured for glass beads contacting aluminum at 1 m/s. The glass beads have a mass flow rate of $m_r = 0.65$ g/s, an average radius of $R = 275$ µm, and a density of $d = 2.5$ g/cm$^3$. Therefore, the charge per contact is approximately $4\pi R^3 dI/3m_r = 1.05 \cdot 10^6$ e.

The magnitudes of $\Delta Q$ from our model compare well to the previous empirical model and the experiment with glass beads. Given the large variability in triboelectric experiments (e.g., the current in [84] ranges from less than 0.2 nA to more than 0.8 nA for 1 m/s contacts) and the rough approximation in applying our model to other materials, these are in reasonable quantitative agreement. As mentioned before, our approach is fully *ab initio* as it requires no adjustable parameters, only the material properties, which is different from other approaches.

## Appendix 7. Calculations with additional effects

There are a number of higher-order electromechanical or contact mechanics effects that we have ignored in the main text. Here, we include quantitative results for some extended cases, and estimate the quantitative impacts of others, as presented below in Table A1.

First, note that we assumed a rigid indenter above. A calculation with an indenter with the mechanical properties of $Pt_{0.8}Ir_{0.2}$, $Y = 230$ GPa and $\nu = 0.37$ [55] is labeled "non-rigid indenter". With a soft indenter, the contact radius is larger and the strains in the semiconductor are smaller, so $\Delta Q_d$ is increased and $\Delta Q_{if}$ and $\Delta Q_{sf}$ are reduced.

Second, a calculation using the JKR adhesive contact theory [91] is labeled "JKR adhesion". The adhesion energy of $Pt/SrTiO_3$ interfaces is about $\Gamma = 1$ J/m$^2$ [123], which adds to the effective force during contact. For the nominal contact in the main text with $F = 4$ μN and $R = 60$ nm, the effective force is $F_{JKR} = F + 3\pi\Gamma R + \sqrt{6\pi\Gamma F R + (3\pi\Gamma R)^2} = 6.75$ μN. The higher force leads to increases in $\Delta Q_d$, $\Delta Q_{if}$, and $\Delta Q_{sf}$ by a factor of $(F_{JKR}/F)^{2/3} \approx 1.4$. Note that this factor depends on the relative values of $\Gamma$, $F$, and $R$.

Third, a calculated labeled "gradient elasticity" uses contact mechanics solutions that incorporate strain gradient elasticity, i.e., the coupling between stress and strain gradients. These calculations use the solutions of Gao and Zhou [66, 67] with the $SrTiO_3$ gradient length scale parameter $\ell = 4$ Å parameter estimated by Stengel [68]. Gradient elasticity regulates the extreme strain gradients that occur near the edge of the contact in standard Hertzian solutions. This reduction in the strain gradient consequently causes a reduction in the flexoelectric terms $\Delta Q_{if}$ and $\Delta Q_{sf}$.

| Theory | $\Delta Q_d$ (e) | $\Delta Q_{if}$ (e) | $\Delta Q_{sf}$ (e) |
|---|---|---|---|
| Main text theory | 189 | 693 | -648 |
| Non-rigid indenter | 306 | 544 | -508 |
| JKR adhesion | 268 | 982 | -918 |
| Gradient elasticity | 189 | 534 | -485 |

Table A2. Comparison of calculation results using the main text theory and various modifications. Note that the modifications are not cumulative, i.e., JKR adhesion and gradient elasticity assume a rigidi indenter.

Beyond the calculations presented in Table A1, which all show changes of less than a factor of 2, there are other effects on the contact mechanics that may be considered. For example, coupling between flexoelectricity and stress [93, 94] and the effects of surface stress [92] are both relevant for these types of contacts. The solutions for these are more complicated and have not been incorporated in our calculations, but they show quantitative changes of roughly a factor of 2 or less when compared to simpler contact theories.